# Hydrazine-Free Precursor for Solution-Processed All-Inorganic Se and Se$_{1-x}$Te$_x$ Photovoltaics

Adam D. Alfieri[1], Swarnendu Das[2], Kim Kisslinger[3], Chloe Leblanc[1], Jamie Ford[4], Cherie R. Kagan[1,2,5], Eric A. Stach[2], Deep Jariwala[1]*

[1]Electrical and Systems Engineering, University of Pennsylvania, Philadelphia, PA, United States, 19104
[2]Materials Science and Engineering, University of Pennsylvania, Philadelphia, PA, United States, 19104
[3]Center for Functional Nanomaterials, Brookhaven National Laboratory, Upton, NY, United States, 11973
[4]Singh Center for Nanotechnology, University of Pennsylvania, Philadelphia, PA, United States, 19104
[5]Chemistry, University of Pennsylvania, Philadelphia, PA, United States, 19104
*Corresponding author: dmj@seas.upenn.edu

## Abstract

Selenium (Se) has reemerged as a promising absorber material for indoor and tandem photovoltaics (PVs), and its alloys with Te (Se$_{1-x}$Te$_x$) offer a widely tunable bandgap. Solution processing of this materials system offers a route to low-cost fabrication. However, solution processing of Se has, thus far, only used hydrazine, which is an extremely hazardous solvent. In this work, we prepare and isolate propylammonium poly-Se and poly-Se-Te precursors from a safer thiol-amine solvent system. We formulate molecular inks by dissolving the precursor in dimethylformamide (DMF) with a monoethanolamine (EA) additive and process high-quality Se and Se$_{1-x}$Te$_x$ films with bandgaps ranging from 1.20 eV to 1.86 eV. We fabricate PVs from these films using TiO$_2$ and MoO$_3$ charge transport layers (CTLs) to achieve power conversion efficiencies as high as 2.73% for Se and 2.33% for Se$_{0.7}$Te$_{0.3}$ under solar simulation. Se devices show excellent stability with no degradation after 1 month in air, enabled by the excellent stability of Se and the use of inorganic CTLs. This work represents an important step towards low-cost solution-phase processing of Se and Se$_{1-x}$Te$_x$ alloys for PVs and photodetectors with low toxicity and high bandgap tunability.

## Introduction

The world's first PV material, trigonal Se[1], has experienced a renaissance as a promising absorber material for both tandem and indoor PVs[2–5], with PCEs of 8.1% achieved under AM1.5G illumination[6] and over 26% PCE under indoor lighting[7], surpassing commercial amorphous-Si:H for this application. Se exhibits strong absorption, good carrier mobility, low toxicity, and a unique combination of low processing temperatures and stability. An added benefit of Se is its broad tunability through isomorphic alloying with Te, enabling the alloy bandgap to be tuned from 0.3 eV to 1.9 eV with just 2 elements and attracting some attention for optoelectronics[8–11]. Typically, Se is processed by thermal evaporation, limiting throughput and increasing costs. Given the low melting point of Se (217°C), melt processing has been successfully pursued[7,12], but this is difficult to control and results in excessively thick (> 2 $\mu$m) films that lead to material waste. Solution processing is an alternative approach that offers significantly higher throughput and lower capital costs than vacuum processing with less material consumption than melt processing, driving down manufacturing costs.

Solution processing of Se PVs has been shown using hydrazine[13,14], which is an effective but extremely hazardous solvent, limiting both widespread research and eventual commercialization using this method. Instead, elemental Se and Te can be dissolved in the thiol-amine solvent system[15], which is far safer than hydrazine. $Se_{0.7}Te_{0.3}$ films for solar cells were prepared from an ethylenediamine:ethanethiol (EDA:ET) solvent at temperatures ≤200°C, with a device efficiency of 1.1%[10]. Thiol-amine processed Se and $Se_{1-x}Te_x$ PVs are therefore worth pursuing for scalable production of stable PVs for indoor and solar applications. However, thiol-amine processing has yet to be applied to pure Se cells, and further development of thiol-amine-based precursors is needed.

Thiol-amine processing of Se and Te has primarily focused on metal selenides and metal tellurides[16]. It was found that dissolving Se in alkylamine:ET (AA:ET) creates an alkylammonium poly-selenide (AAPSe) species that acts as an "alkahest"[17] capable of dissolving a variety of metals. The AAPSe alkahest was shown to be able to dissolve Te, which is insoluble in AA:ET solutions in the absence of Se[18]. Changing the amine (AA vs EDA) impacts the precursor. Given the importance of precursor selection to the morphology and performance of solution processed thin film PVs, identifying a precursor most suitable for producing films and then developing a process based on that precursor is a crucial first step towards solution processed Se and $Se_{1-x}Te_x$ PVs.

In this work, we use precursor and process engineering to control the morphology of Se and $Se_{1-x}Te_x$ films produced by thiol-amine processing. We demonstrate that propylammonium poly-selenide (PAPSe) and poly-$Se_{1-x}Te_x$ (PAPST) salts prepared from propylamine (PA) and ET can be redissolved in benign solvents and spin-coated on preheated substrates to achieve uniform, high-quality films of Se and $Se_{1-x}Te_x$ alloys with widely tunable bandgaps. We fabricate and characterize photovoltaics from the films and achieve a respectable PCE of 2.73% and open circuit voltage ($V_{oc}$) as high as 854 mV for Se, comparable to that of evaporated films. Similarly, we achieve 2.33% PCE for ~1.20 eV bandgap $Se_{0.7}Te_{0.3}$ films, which is more than double the previous report on solution processed $Se_{1-x}Te_x$ PVs.

**Results**
*Precursor*
We prepare two different precursors for pure Se, shown in Figure 1, through the dissolution of elemental Se in two thiol-amine solvent systems. First, we consider dissolving Se in a PA/ET solvent system. This results in "dimensional reduction" of the Se chain to form short poly-selenide chains coordinated by propylammonium ions, similar to the Se species present when dissolved in hydrazine[17,19]. Deprotonated thiols react with each other to form diethyl disulfide (DEDS). Toluene is added to precipitate the propylammonium poly-selenide (PAPSe) salt. Toluene, DEDS, and excess unreacted amine/thiol are extracted by decanting the supernatant and vacuum drying. The solid PAPSe salt can then be dissolved in polar aprotic solvents. Alternatively, elemental Se is dissolved in EDA:ET, producing an ethanethio-poly-selenide (ETPSe) anion coordinated by the $H_2NC_2H_4NH_3^+$ cation in which the ethanethio- ligand is bonded to the poly-selenide chain. Here, excess EDA and ET are similarly extracted using toluene and vacuum, but the resulting compound is a viscous liquid that can be diluted with EDA or polar solvents. Both PAPSe and ETPSe can be converted to trigonal Se upon annealing (Figure S1). Differences between diamine-thiol and monoamine-thiol precursors are discussed in more detail in another previous work[18].

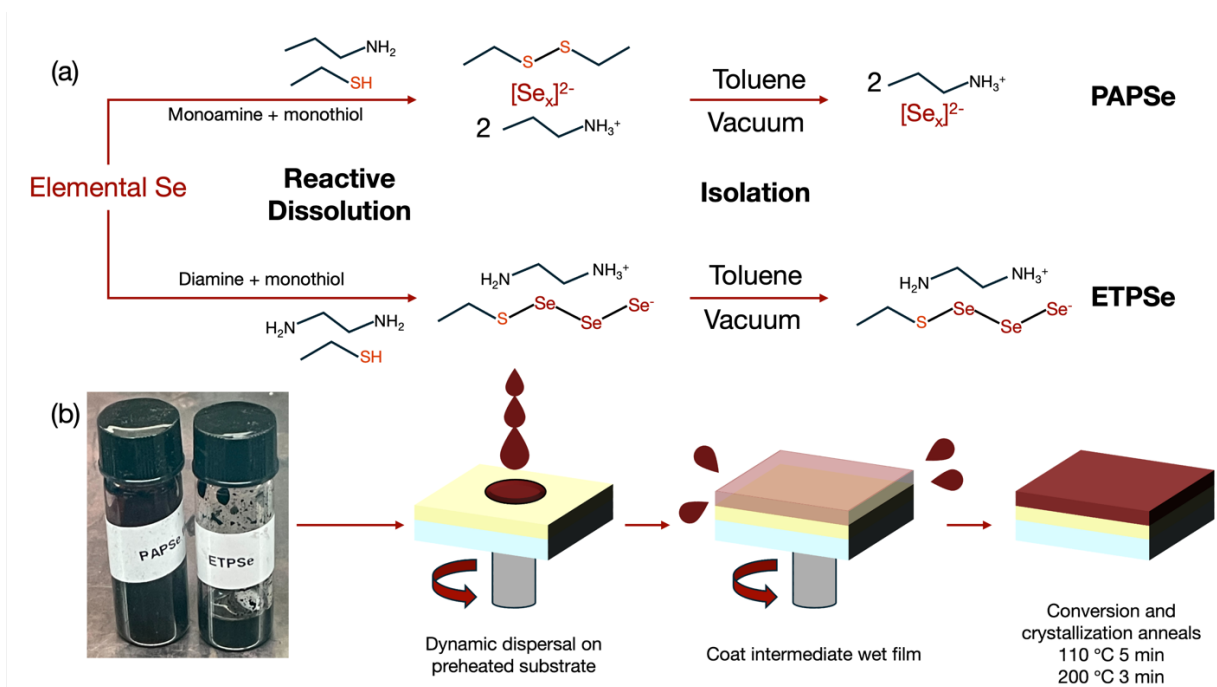

**Figure 1. Precursor chemistry and deposition process**. (a) Dissolution of elemental Se in propylamine and ethanethiol results in a different precursor than dissolution in ethylenediamine and ethanethiol. Preparation of Se$_{1-x}$Te$_x$ alloys follows the same precursor form. (b) Images of PAPSe and ETPSe precursor inks and schematic of deposition process.

We make films by dissolving the PAPSe salt in DMF with a small (2% by volume) EA additive. The EA additive was found to increase grain size and improve device performance (Figure S2), presumably due to the high boiling point and interaction with Se. Similarly, we dilute ETPSe precursor to approximately 4M with DMF. The films are deposited by dynamic spin-coating of the solution on mesoporous- (mp-)TiO$_2$/compact- (c-)TiO$_2$/FTO/glass substrates preheated to 110°C, converting to Se by annealing at 110°C, and then crystallizing the films at 200°C (see Supplementary Information for Methods). We found that preheating the substrate to 110°C resulted in greater reproducibility and film adhesion/uniformity compared to an unheated substrate and/or using an antisolvent, though antisolvent engineering could be a future path to control film nucleation and growth. Figure 2a shows the camera image (top row) and corresponding SEM image (bottom row) of a representative Se film from ETPSe, and Figure 2b shows the same for the Se film from PAPSe. We were unable to achieve uniform films from ETPSe, as the precursor tends to form large discontinuous clusters on top of an infiltrated mp-TiO$_2$ layer instead of films. In contrast, PAPSe films are continuous, uniform, and have well defined grains with sizes on the order of several hundreds of nanometers, comparable to evaporated Se films with similar thicknesses[3]. The morphology is dense, except for small pinholes to the mp-TiO$_2$ that will need to be eliminated with process optimization. Overall, Se films from the PAPSe precursor clearly have a wider processing window than Se films from ETPSe, and we focus on PAPSe-processed films for the remainder of this work.

*Alloying with Te*

We also use the PA-based precursor to make $Se_{1-x}Te_x$ alloys using the same precursor chemistry by adding stoichiometric amounts of Se and Te powder during the precursor preparation, producing propylammonium poly-seleno-telluride (PAPST) salts. Precursors with Te concentrations of 10% and 20% Te were used, resulting in films with compositions of approximately 14% Te ($Se_{0.86}Te_{0.14}$) and 30% Te ($Se_{0.7}Te_{0.3}$), estimated by energy dispersive X-ray spectroscopy (EDX, Methods). Se loss is likely due to a combination of higher solubility of Se-rich precursor in the precursor ink and the higher vapor pressure of Se resulting in Se loss during annealing[10]. Cross-sectional scanning transmission electron microscopy (X-STEM) EDX analysis of the 30% Te film confirms the homogenous alloying of Se and Te and complete infiltration of the mp-$TiO_2$ layer (Figure S3). Camera and SEM images of the 14% and 30% Te films are shown in Figure 2c-d. Adding Te results in strongly absorbing, smooth films (RMS roughness of 10.0 nm for 14% Te, 5.7 nm RMS roughness for 30% Te, Figure S8) with significantly smaller grains (tens of nm) that are difficult to resolve. The decrease in grain size is consistent with previous observations[9] and the hypothesis that adding Te requires short chains[20]. The films produced from PAPST nonetheless highlight the potential and flexibility of this precursor system.

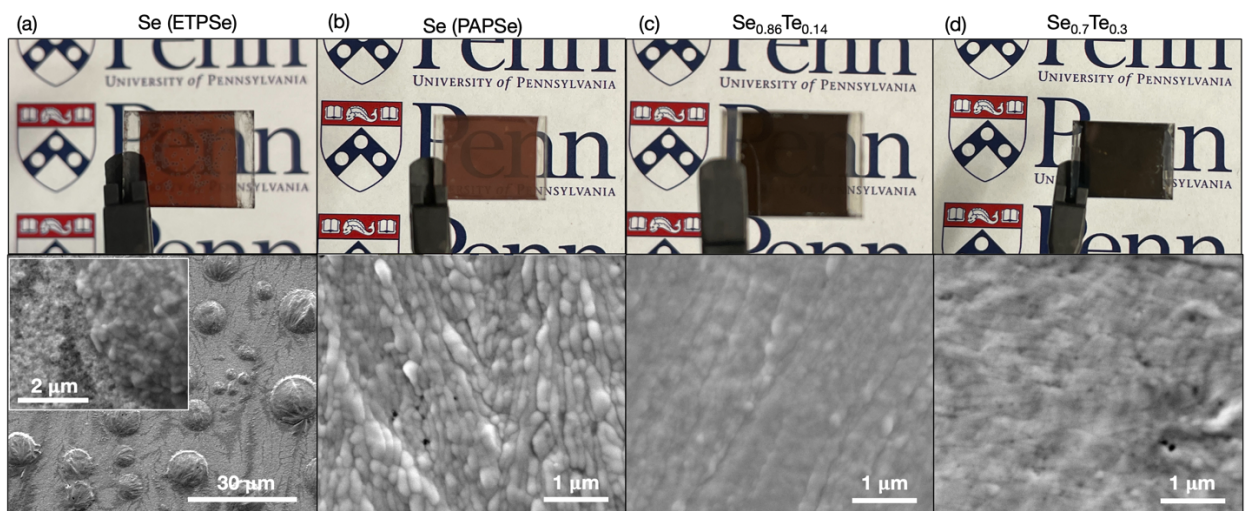

**Figure 2. Optical and Electron Microscope Images of Films.** Camera (top row) and SEM (bottom row) images of (a) Se processed from ETPSe, (b) Se processed from PAPSe, (c) $Se_{0.86}Te_{0.14}$, and (d) $Se_{0.7}Te_{0.3}$ films. The camera images are taken from the bottom side of the 20 mm x 15 mm $TiO_2$/FTO/glass substrate.

The bandgap as a function of Te concentration can be estimated using UV-Vis absorption spectroscopy. The absorbance for the various concentrations is plotted in Fig. 3a. The estimated bandgaps from fitting the absorbance are 1.83 eV for pure Se, 1.54 eV for 14% Te, and 1.25 eV for 30% Te, which are consistent with the bandgap determined from device external quantum efficiency (EQE) measurements: 1.86 eV, 1.48 eV, and 1.20 eV (Figure 4e). This wide range of bandgap tunability using solution processing with just 2 elements makes this system attractive for indoor PVs, tandem top cells, and single junction PVs.

Raman spectra for the 3 samples is shown in Figure 3b. The sharp peak at 234 cm$^{-1}$ is consistent with the trigonal phase, and the absence of a peak at 251 cm$^{-1}$ for Se indicates the lack of an amorphous phase. The peak at 234 cm$^{-1}$ can be deconvoluted into nearly degenerate $E$ and $A_1$ modes, hence the shoulder on the peak. As Te is added, peaks get weaker and broader, redshift, and split into multiple peaks with the emergence of new modes[21]. The composition-dependent positions of the peaks the alloy films are consistent with previous reports[21], indicating that the compositions extracted from EDX are reasonable estimates despite the inherent uncertainty associated with EDX.

The XRD data (Figure 3c) shows (100) and (101) diffraction peaks for Se, consistent with the desired trigonal phase. The ratio of the desired (101) peak at ~29.7° to the (100) peak at ~23.6° is comparable or superior to films produced by evaporation of Se with a Te interlayer on an unheated substrate. Further optimization of deposition conditions to achieve a fully (101) textured film will be beneficial for charge transport and device performance, and this should be a focus of future studies. The 14% Te film shows significantly smaller peaks, indicative of poor crystallinity. Interestingly, as the Te concentration increases to 30%, the crystallinity improves, but with an undesired (100) texture. The shifting of the peaks to lower angles with Te addition is consistent with previous results[10]. The large differences in the XRD spectra with varying Te content show that Te has a large effect on the nucleation and growth kinetics.

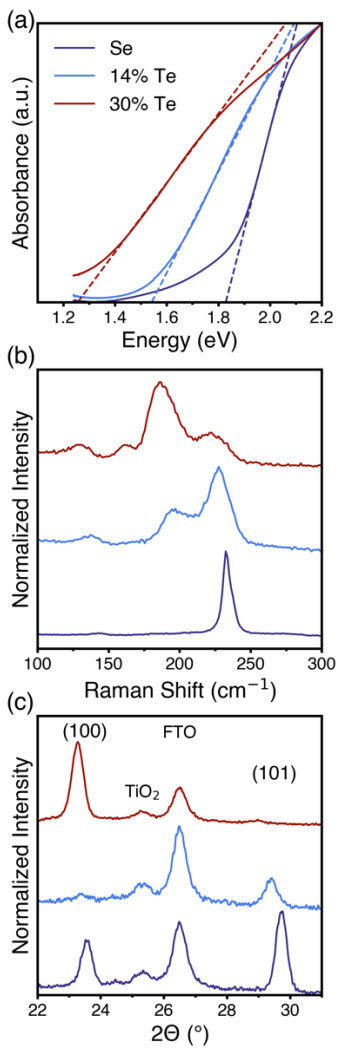

**Figure 3. Film characterization.** (a) UV-Vis absorbance (b) Raman and (c) XRD for varying Te concentration. For this figure and those that follow, the dark blue curve corresponds to pure Se, the light blue corresponds to 14% Te, and the red curve corresponds to 30% Te.

*Device performance*

Devices are fabricated from the PAPST $Se_{1-x}Te_x$ films using the device structure in Fig. 4a: FTO/c-$TiO_2$/mp-$TiO_2$/$Se_{1-x}Te_x$/$MoO_x$/Au, where the $TiO_2$ is the electron transport layer (ETL) and the $MoO_x$ layer is the hole transport layer (HTL). We note that the ability to use the inorganic, evaporated 15 nm $MoO_x$ HTL is enabled by the uniformity of our films. Using $MoO_x$ instead of doped PTAA or spiro-OMeTAD (as in other prior works[10,13]) will enable significantly greater long-term stability and lower costs. Figure 4b shows a cross-sectional SEM image indicating that the film thickness is approximately 285 nm for Se. The $Se_{1-x}Te_x$ films are approximately 200 nm thick (Figure S4).

Fig. 4c shows the J-V curves under simulated AM1.5G light (intensity of 93.4 mW/cm$^2$) with an LED based solar simulator (Methods). Figure 4f shows the statistics for the device efficiency of Se and alloy cells; full device statistics (including $J_{sc}$, $V_{oc}$, and FF) are provided in Figure S5, and the maximum and mean values of $J_{sc}$, $V_{oc}$, FF, and PCE are listed in Table 1. Pure Se cells initially showed a PCE as high as 2.48%. After aging the devices in air (in the dark) and re-measuring, the PCE improves to 2.73%. The improvement after aging is a well-known phenomenon for devices with a $MoO_x$ HTL. While the overall efficiency is lower than evaporated Se PVs and one report of hydrazine-processed Se PVs, our Se cells exhibit an open circuit voltage ($V_{oc}$) as high as 854 mV, ~200 mV better than cells processed with hydrazine and only ~100 mV less than the best evaporated Se PVs. However, our devices exhibit significantly lower short circuit current density ($J_{sc}$), ultimately resulting in worse efficiencies. The short circuit current density is comparable to Te-free interfaces in evaporated cells[4], suggesting that the interface could be a limiting factor. C-V analysis (Figure S6) indicates that the defect density is in the range of $10^{18}$ cm$^{-3}$. However, this is almost entirely due to interface states, as drive-level capacitance measurements – which minimize interfacial contributions[22] – give bulk defect densities as low as 5.9x10$^{16}$ cm$^{-3}$ (Figure S6), which is only one order of magnitude higher than state-of-the-art evaporated cells[23]. Despite the efficiency deficit, the comparable $V_{oc}$ of PAPSe-processed Se photovoltaics to evaporated films suggests that PAPSe-processed Se devices have the potential to reach similar or even improved efficiencies with greater understanding and optimization of the film processing and the buried interface, which should be a strong focus for future research and development in the community.

The devices from alloyed films show a drastic decrease in $V_{oc}$ and FF, but the $J_{sc}$ is enhanced, particularly for 30% Te, which approaches a $J_{sc}$ of 20 mA/cm$^2$ in a 200 nm thick film. This trend has been previously seen in evaporated $Se_{1-x}Te_x$ PVs[9]. The best 30% and 14% devices have PCEs of 2.33% and 1.70%, respectively. There is significant difference in the performance between 14% and 30% Te despite the bandgaps, 1.5 eV and 1.2 eV, both being in the approximately optimal range for single junction PVs, which we assume to be related to the improved crystallinity of the 30% Te film. This suggests that there is significant room for optimization of single junction $Se_{1-x}Te_x$ PVs by composition within the ideal bandgap window. The champion alloy PCE 2.33% doubles the efficiency of the previous report on solution processed $Se_{1-x}Te_x$ cells[10], and is approaching the efficiency of the best evaporated $Se_{1-x}Te_x$ cells[9]. This value is comparable to other novel lead-free technologies and not especially impressive, but the bandgap tunability, low toxicity, and facile low-temperature processing makes the $Se_{1-x}Te_x$ system worth further exploration. It is also worth mentioning that while Te is a scarce element, a ~200 nm thick (including the $TiO_2$ scaffold), 30% Te absorber could still be more sustainable than other established thin film PV technologies with scarce elements and thicknesses of several microns, e.g. Cu(In,Ga)S$_2$ and CdTe.

Figure 4d shows the EQE spectra of representative devices with varying Te concentration. EQE values reach a maximum of approximately 47% for pure Se. The drop in EQE at shorter wavelengths may be the result of interfacial recombination. The addition of Te extends the spectral response, as can be expected from the UV-Vis data. The 30% Te spectrum achieves maximum EQE values approaching 60% in the visible regime and exhibits a spectral response extending beyond 900 nm. The EQE begins to drop at wavelengths of 700 nm and longer, which is possibly the result of either incomplete absorption in this range or poor long wavelength photon collection due to recombination in the space-charge region. The bandgaps are extracted from -ln(1-EQE) (Figure 4e) to be 1.86 eV, 1.48 eV, and 1.20 eV for pure Se, 14% Te, and 30% Te. The value for pure Se is consistent with previous estimates, and the high bandgap tunability through Te alloying is again apparent.

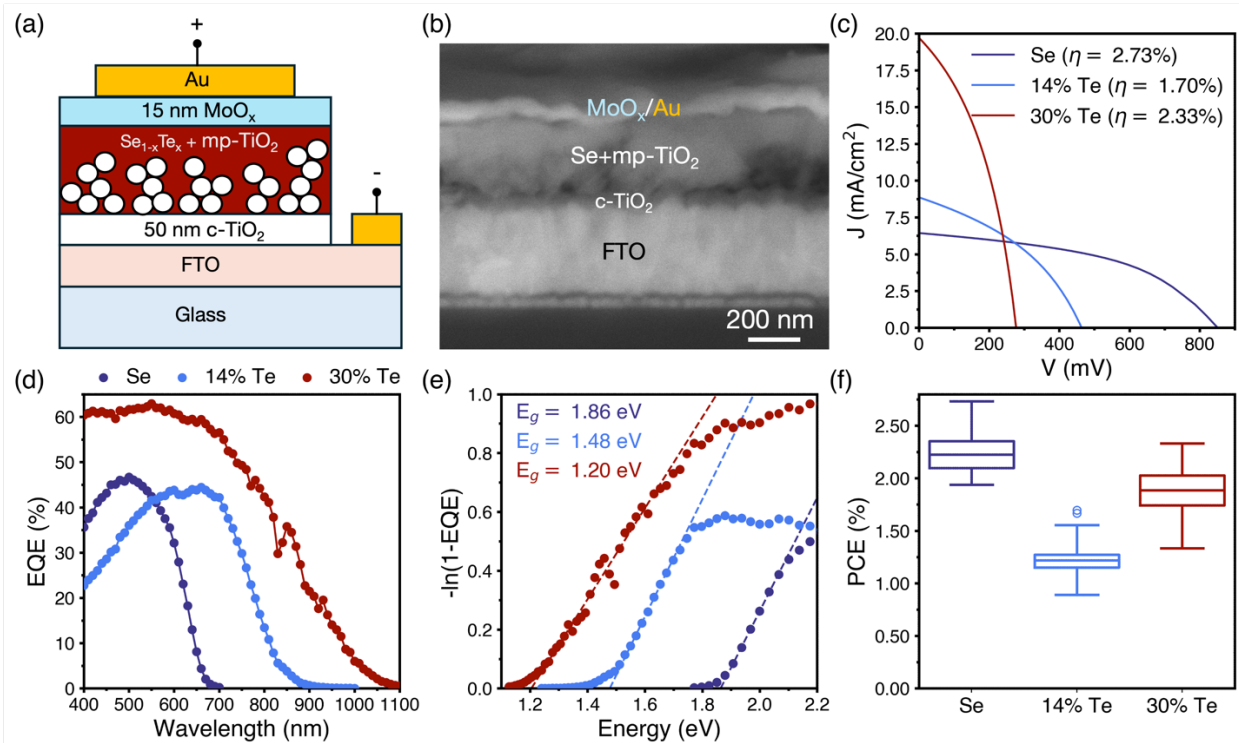

**Figure 4. Photovoltaic device performance.** (a) Device structure schematic and (b) cross sectional SEM image of a Se device. Devices have a mesoscopic superstrate configuration with a TiO$_2$ electron transport layer and a MoO$_x$ hole transport layer. (c) J-V curves of devices from different Te concentrations. (d) EQE spectra and (e) -ln(1-EQE) spectra for different Te concentrations with bandgaps extracted and listed. (f) Boxplot of the PCE as a function of Te concentration. To account for device aging, the J-V scan with the highest PCE from each device over time is included in the boxplot.

**Table 1.** Champion (mean) device characteristics for PVs from Se and various Te concentrations.

| Material | $J_{sc}$ (mA/cm$^2$) | $V_{oc}$ (mV) | FF (%) | PCE (%) |
|---|---|---|---|---|
| **Se** | 6.6 (5.7) | 854 (824) | 50.1 (47.3) | 2.73 (2.24) |
| **14% Te** | 8.8 (6.8) | 502 (445) | 42.7 (40.4) | 1.70 (1.24) |
| **30% Te** | 19.8 (16.4) | 304 (269) | 44.1 (42.2) | 2.33 (1.87) |

The dark J-V characteristics of the devices are plotted on a semi-log scale in Figure 5a. Se is known to be highly resistive when not illuminated, which is reflected in the J-V curve. Adding Te increases the current density under forward bias but also reduces the shunt resistance for 30% Te, which is detrimental to device performance. The high shunt conductance could result from Te-rich regions and/or the low (0.27V) built-in potential. The high shunt conductance is consistent with previous work[9,10].

We perform illumination intensity-dependent J-V measurements of representative devices and extract the ideality factor using the illumination dependent $V_{oc}$ method (Figure 5b). The Se devices have an ideality factor ~2, indicating Shockley-Read-Hall (SRH) recombination dominates. It is also possible that the higher ideality factor and slight nonlinearity for Se is related to previously reported illumination-dependent mobility[24], though this will become clearer with improved film and interface quality. Devices from both alloys exhibit an ideality factor below 2, indicating a mixture of bimolecular and SRH recombination mechanisms. The greater defect density of 14% Te films (Figure S6) likely results in a higher ideality factor (1.60) than 30% Te (1.31). The trend of the ideality factors (decreasing with increasing Te content) is consistent with the slope of the linear region of the semi-log dark J-V curves in Figure 5a.

The low $V_{oc}$ and FF of $Se_{1-x}Te_x$ alloys is likely attributable to a combination of higher defect densities and low built in-voltages (Figure S6). 30% Te cells have comparable (~$4 \times 10^{16}$) bulk defect densities to pure Se, but more interface states/doping. 14% Te cells have on the order of $2 \times 10^{17}$ cm$^{-3}$ bulk defects and $2 \times 10^{19}$ cm$^{-3}$ total defects, hence the poor performance. Te point defects in Se are benign[25], so it is likely that the significant decrease in grain size introduces the higher defect densities. This would be consistent with a previous study indicating that Te alloying results in extended defects acting as electron traps[26]. The built-in potentials, 0.52V and 0.27V for 14% and 30% Te, respectively, are similar to the $V_{oc}$ for these materials, indicating that engineering the $V_{bi}$ could be necessary to aid charge separation and suppress dark current.

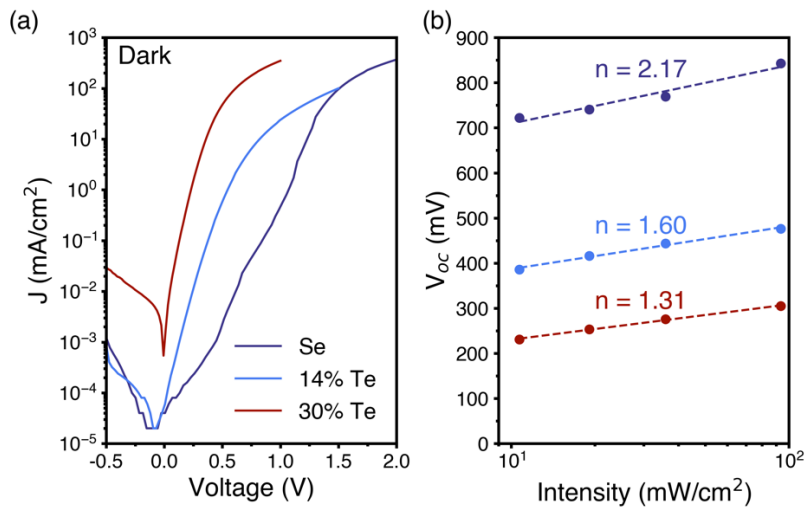

**Figure 5. Dark and Suns-$V_{oc}$ Characterization as a function of Te concentration. (**a) Semi log dark J-V characteristics. (b) Illumination-dependent $V_{oc}$ and ideality factor fit.

To characterize the stability, unencapsulated devices are stored in the dark in ambient air and repeatedly measured over time. Figure 6a shows the PCE of representative devices over time. Se cells exhibit an improved PCE after aging for 1-2 weeks and show no performance degradation after over 1 month. The improvement after aging is a well-known phenomenon for devices with a $MoO_x$ HTL due to healing of oxygen vacancies and has previously been observed for Se cells[27]. While accelerated aging tests will eventually be needed for Se cells once they become more commercially feasible, our results show the benefit of Se's air stability. In contrast, the performance of $Se_{1-x}Te_x$ cells degrades with time, with greater degradation for higher Te content. Cells stored in $N_2$ do not show significant degradation for 30% Te and show improvement in the 14% Te cells (Figure 6b), so the degradation mechanism can be attributed to Te oxidation. These results suggest $Se_{1-x}Te_x$ solar cells have the potential to be a stable technology, but alloys will likely require encapsulation.

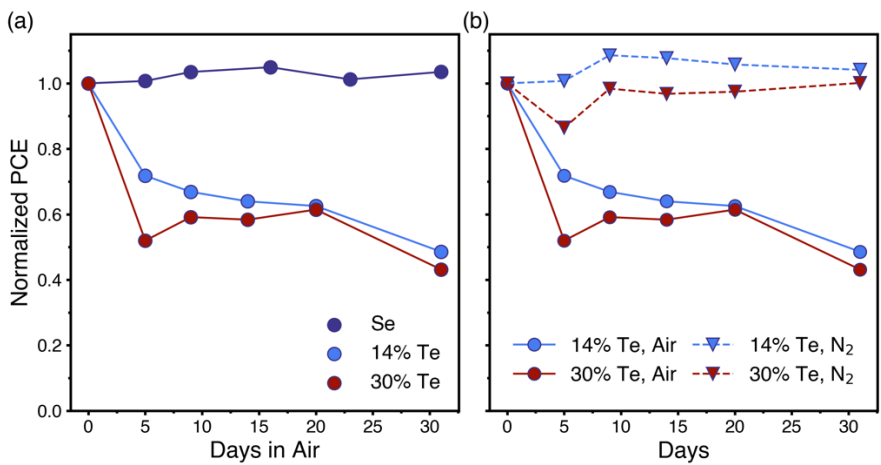

**Figure 6. Stability of Unencapsulated Devices.** (a) Normalized PCE over time for devices stored in dark in air and remeasured over time. (b) Comparing the stability of $Se_{1-x}Te_x$ cells stored in air vs stored in an $N_2$-filled glovebox over time. All devices are measured in air.

**Outlook and conclusions**

Continued work to understand and control the crystallization process will be necessary to produce improved Se films with reduced defect densities. It is particularly important to understand intermediate phases and pathways between the PAPSe precursor and trigonal Se to tailor the ink, deposition, and annealing process. Understanding how to control the film formation with temperature, solvents, and additives will yield better device performance, but developing this understanding is a long process that is still ongoing for lead-halide perovskites after over a decade of research, for example. Work can also be done to improve the PAPSe precursor itself through optimized PA:ET:Se ratios and better precursor purification. Further, our results suggest that the $TiO_2$/Se interface is a limiting factor. Understanding the interactions between PAPSe and the $TiO_2$ and identifying an interfacial modification strategy could be fruitful in improving performance. For $Se_{1-x}Te_x$ alloys, it will additionally be necessary to control compositional order, which may require changes to the ink formulation or annealing processes. The Te composition has a large impact on both bandgap and microstructure, meaning that the Te composition must be carefully tuned to optimize performance.

In summary, we have developed propylammonium poly-[$Se_{1-x}Te_x$]$_y^{2-}$ (PAPST) molecular inks as a promising precursor for hydrazine-free solution processing of Se and $Se_{1-x}Te_x$ thin films for photovoltaics. We have shown that thiol-amine solution processing can achieve comparable PCE for Se films (2.73%) to those processed by hydrazine, with higher open circuit voltage (up to 854 mV). Similarly, we achieve PCEs for $Se_{0.7}Te_{0.3}$ films of 2.33%, better than previous work on solution processed $Se_{0.7}Te_{0.3}$ films[10] and comparable to evaporated films[8,9]. Using PAPSe/PAPST precursors, neat films with a wide processing window can be achieved for widely tunable bandgaps (1.86 eV to 1.20 eV and possibly lower), highlighting the potential of this processing method and the $Se_{1-x}Te_x$ materials system for next-generation PVs and other optoelectronics.

**Acknowledgements**
D.J. and A.D.A. acknowledge primary support from the University of Pennsylvania Materials Research Science and Engineering Center (MRSEC) DMR-2309043 Seed program and partial support from the Vagelos Institute for Energy Science and Technology. C.R.K acknowledges support from IMOD STC DMR-2019444. This work was conducted in its majority at the Singh Center for Nanotechnology at the University of Pennsylvania, which is supported by the NSF National Nanotechnology Coordinated Infrastructure Program grant no. NNCI1542153. The authors gratefully acknowledge the use of facilities and instrumentation (G2V Pico solar simulator and Cary UV-Vis Spectrophotometer) supported by the Department of Materials Science and Engineering Departmental Laboratory at the University of Pennsylvania. The authors acknowledge the use of an XRD facility supported by the Laboratory for Research on the Structure of Matter and the NSF through the University of Pennsylvania Materials Research Science and Engineering Center (MRSEC) DMR-2309043. This research used Electron Microscopy facilities of the Center for Functional Nanomaterials (CFN), which is a U.S. Department of Energy Office of Science User Facility, at Brookhaven National Laboratory under Contract No. DE-SC0012704.

**Supplementary Information contains Methods and Figures S1-S8.**

# Supplementary Information: Hydrazine-free Precursor for Solution-Processed All-Inorganic Se and Se$_{1-x}$Te$_x$ Photovoltaics

Adam D. Alfieri[1], Swarnendu Das[2], Kim Kisslinger[3], Chloe Leblanc[1], Jamie Ford[4], Cherie R. Kagan[1,2,5], Eric A. Stach[2], Deep Jariwala[1]*

[1]Electrical and Systems Engineering, University of Pennsylvania, Philadelphia, PA, United States, 19104
[2]Materials Science and Engineering, University of Pennsylvania, Philadelphia, PA, United States, 19104
[3]Center for Functional Nanomaterials, Brookhaven National Laboratory, Upton, NY, United States, 11973
[4]Singh Center for Nanotechnology, University of Pennsylvania, Philadelphia, PA, United States, 19104
[5]Chemistry, University of Pennsylvania, Philadelphia, PA, United States, 19104
*Corresponding author: dmj@seas.upenn.edu

## Methods

*Materials*
Fluorinated tin oxide (FTO, TEC 15, on 20 mm x15 mm x2.2 mm soda lime glass) substrates are purchased from Ossila. Black TiO$_{2-x}$ evaporation pellets (99.9%) are purchased from Kurt J. Lesker. NRD-30 TiO$_2$ paste is purchased from Greatcell Solar. Selenium powder (99.999%, 200 mesh), Tellurium powder (99.999%, 200 mesh), 2-methoxyethanol (2-ME, 99.9%) ethanethiol (ET, 99+%), ethylenediamine (EDA, 99+%), monoethanolamine (EA, 99+%), and n,n-dimethylformamide (DMF, anhydrous, extra dry, 99.9%) are purchased from Thermo Fisher Scientific. *n*-propylamine (PA, 99+%) and toluene (anhydrous, 99.8%) are purchased from Sigma Aldrich. Electronic grade hydrogen peroxide (H$_2$O$_2$), ammonium hydroxide solution (NH$_4$OH), isopropyl alcohol (IPA), and acetone are from Transene.

*Substrate Preparation*
FTO substrates are sonicated in detergent solution, DI water, 75°C acetone and IPA for 10 min each and dried with N$_2$. Next, the substrates are cleaned in 1:1:6 NH$_4$OH:H$_2$O$_2$:DI water at 80°C for 10 min before thoroughly rinsing with DI water and drying with N$_2$. The substrates are treated with a 50W O$_2$ plasma for 4 min before a 50 nm compact TiO$_2$ layer is deposited by electron beam (E-beam) evaporation at a rate of 0.5 Å/s and a base pressure of < 3.5x10$^{-6}$ torr. The substrate is treated with a 50W remote Ar+O$_2$ plasma for 2 min to make the surface hydrophilic. The mesoporous TiO$_2$ layer is deposited by spin-coating TiO$_2$ paste diluted 1:5 paste:2-ME by weight at 4k rpm before drying at 150°C for 10 min and sintering at 510°C for 30 min on a hotplate in air. The paste:2-ME solution is filtered through 0.22$\mu$m nylon syringe filters and then sonicated while heating at 40°C until immediately before deposition to prevent agglomeration. Prior to deposition of Se and Se$_{1-x}$Te$_x$ films, the substrates are again treated with a 50W remote Ar+O$_2$ plasma for 2 min.

*Molecular Ink Preparation from EDA:Thiol*
We prepare ETPSe molecular ink by dissolving Se powder in 2.5 mL of 4:1 EDA:ET at a Se concentration of 6M, its maximum solubility. The mixtures are magnetically stirred at 300 rpm for 1 h at room temperature before filtering through 0.45 um PTFE syringe filters to remove any particulates or undissolved material. We then added 2 mL toluene antisolvent to the solution and centrifuged for 5 min. This resulted in a translucent supernatant on top of a black ink. The supernatant is then decanted. Next, the solutions are subjected to a vacuum of *ca.* 1 bar at room temperature for 15 min to evaporate remaining volatile components (excess EDA, ET, toluene). This results in viscous, concentrated solutions that are then diluted to concentrations of 2-4 M with DMF, EDA, EA, or combinations thereof. For the film shown in Figure 2a, the film was processed from a 3M ink diluted with pure DMF.

*Propylammonium poly-Selenotelluride (PAPST) Precursor Solution Preparation*
Stoichiometric amounts of Se and Te powder are dissolved in 2 mL of 1:1 PA:ET at total chalcogen concentrations of 2M-4M, depending on the Te content. The solubility is higher for lower Te content and pure Se. The mixtures are magnetically stirred at 300 rpm for *ca.* 1 h for pure Se and longer (up to 24 h) for alloys. The PAPST salt is then precipitated in excess toluene and centrifuged for 10 min before decanting the supernatant and drying under vacuum for 30 min.

*NOTE: Ethanethiol reacts with humid air to form toxic $H_2S$ gas! Use of ethanethiol is done in a glovebox, and centrifuge vials containing ethanethiol should be kept tightly sealed when removed from the glovebox.*

For pure Se, the isolated PAPSe salt is dissolved in DMF with 2% EA by volume at a concentration of 465 mg/mL. For PAPST with 10% and 20% Te, the EA content is increased to 4% EA and the concentration is 400 mg/mL for 10% Te and 350 mg/mL for 20% Te, respectively. The exact molarity is unknown because of Se and Te loss during the isolation step and the variation of the PAPST molar mass, hence the units of mg/mL for concentration. After aging the solution for approximately 1 h to ensure full dissolution, the solution is filtered through 0.45 um PTFE syringe filters to remove any particulates.

*Film Deposition*
Prior to absorber deposition, mp-$TiO_2$/c-$TiO_2$/FTO substrates are preheated on a hotplate set to 110°C. The absorber layer is deposited in a glovebox by spin-coating using a single step of 3k rpm for 40s with 125 $\mu$L of precursor solution dynamically dispensed at ~5s. Afterwards, the film is annealed on a hotplate at 110°C for 5 min in a glovebox before a higher temperature anneal of 200°C for 3 min in ambient air on a hotplate. The temperatures listed are the surface temperature of the hotplate, measured by an infrared thermometer.

*Film Characterization*
UV-Vis-NIR measurements are performed using an Agilent Cary 5000 Spectrophotometer from 350 to 1000 nm. Raman measurements are performed using a Horiba LabRam confocal microscope with a 633 nm excitation laser. X-ray diffraction (XRD) measurements are performed using a Rigaku Smartlab diffractometer. Scanning electron microscope (SEM) images are collected with a FEI Quanta 600 ESEM at an accelerating voltage of 5 kV. Composition of the alloys is estimated using energy dispersive X-ray spectroscopy (EDX) at using the same FEI Quanta ESEM at an accelerating voltage of 30 kV with an EDAX detector and fitting with EDAX TEAM software. The quantitative fitting considers the Te L X-rays and the Se K X-rays. RMS roughness is determined by atomic force microscopy (AFM, Figure S8), which is performed using an AIST-NT SmartSPM-1000.

*Device Fabrication*

After absorber layer deposition, 15 nm molybdenum trioxide (MoO$_3$) powder is thermally evaporated at a rate ca. 0.5 Å/s with a base pressure < 3.5x10$^{-6}$ torr. The stacks are then annealed on 180°C in air for 2 min (Se) or 10 min (Se$_{1-x}$Te$_x$). Finally, an E-beam evaporator is used to evaporate 100 nm Au through a shadow mask (active area of 0.05 cm$^2$) at a rate of 2 Å/s and a base pressure <10$^{-6}$ torr. Finally, the Se devices are annealed in air at 210°C for 2 min, which enhanced J$_{sc}$ slightly (Figure S7); Se$_{1-x}$Te$_x$ devices performed better without this final anneal.

*Device Characterization*

Solar simulation I-V curves are taken using a G2V Pico LED light source and a Keithley 2400 source meter in ambient conditions at a scan rate of ~0.6 V/s from +1.1V$_{oc}$ to -1.1V$_{oc}$. Dark I-V curves are performed using the same Keithley 2400 source meter. Capacitance measurements (Figure S6) are performed in air at room temperature under dark conditions using a Keithley 4200-SCS parameter analyzer at a frequency of 10 kHz. C-V measurements (30 mV RMS AC amplitude) from -0.5V to 1V for Se cells, -0.5V to 0.8V for 14% Te, and -0.3V to 0.5V for 30 % Te. Drive level capacitance measurements are performed using a 10 kHz AC signal with peak-to-peak AC amplitudes ranging from 15 mV to 135 mV with DC bias adjustment to maintain constant total applied bias. This is performed at DC biases of 0 to 0.8V (Se), -0.1 to 0.4V (14% Te), and -0.2 to 0.2V (30% Te). EQE measurements are performed using a collimated Xe arc lamp light source (Sciencetech) operating at 265W filtered with a monochromator (Newport CS-130B) configured with a spectral resolution of 5 nm. A 600 nm long pass filter (Edmund Optics) is used at the output of the monochromator for wavelengths of 610 nm and longer to filter out higher order diffractions. The optical power density is measured using a calibrated Si photodiode power meter (Thorlabs PM-121) and the photocurrent at zero bias is measured using a Keithley 2450 sourcemeter. We are limited to a spectral range of 400-1100 nm due to the power meter, preventing us from estimating the integrated J$_{sc}$ from EQE measurements due to the inability to accurately determine the contributions from the UV regime. Samples for cross section SEM are prepared by cleaving devices with a glass breaking tool, and imaging is performed using a JEOL 7500F HRSEM with an accelerating voltage of 15 kV. Cross-sectional TEM samples were prepared using the in-situ lift-out method on the FEI Helios G5 UX DualBeam Ga$^+$ FIB/SEM, and progressively thinned using lower beam voltages, with final milling performed at 2 keV. High-angle annular dark-field-scanning transmission electron microscopy (HAADF-STEM) images and the corresponding energy-dispersive X-ray spectroscopy (EDX) maps were acquired using an aberration-corrected JEOL NEOARM equipped with 2 solid state angle detectors operating at 200 kV. The HAADF-STEM images and the associated EDX spectra were recorded using a 4 cm camera length. HAADF-STEM images are acquired using 12 μs per pixel, while EDX spectra were recorded using 0.01 s per pixel with pixel size of 1 nm by 1 nm.

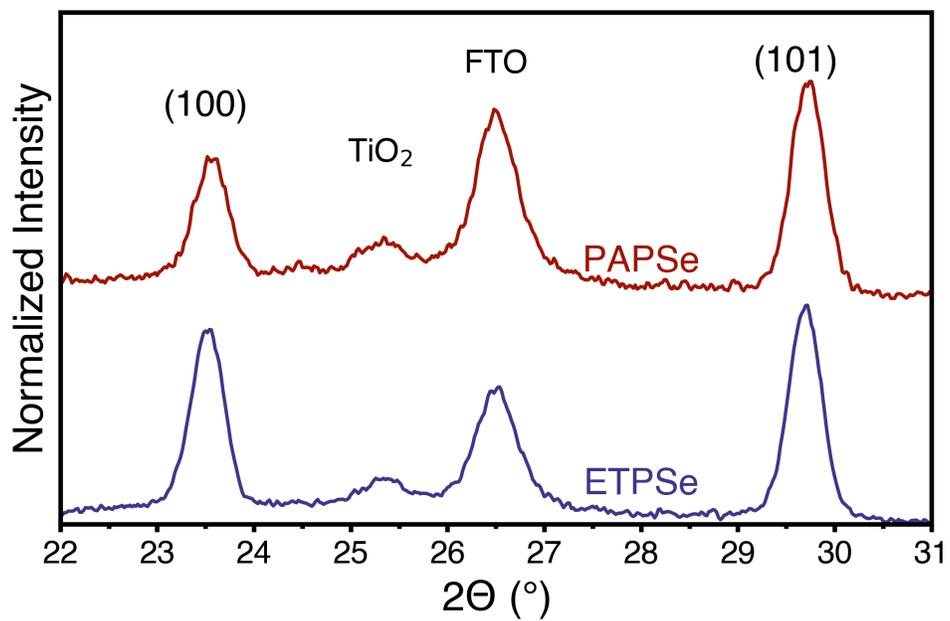

**Figure S1.** X-Ray Diffraction Patterns of Se from PAPSe and ETPSe. Both precursors form trigonal Se.

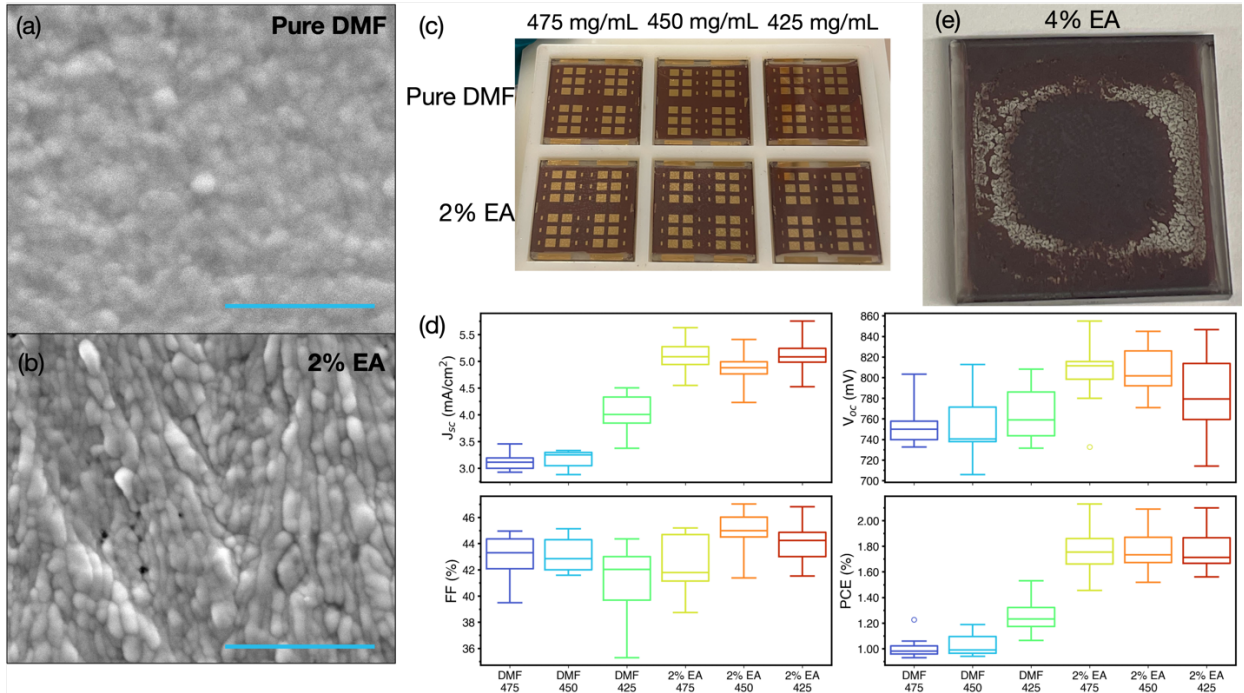

**Figure S2. Effects of EA additive.** SEM images of Se films processed with (a) Pure DMF and (b) DMF with 2% EA. EA films have more distinct, larger grains. The blue scale bars are 2 $\mu$m. (c) Photograph of preliminary devices fabricated from DMF (top row) and DMF+EA (bottom row) with different ink PAPSe concentrations of (left to right) 475, 450, and 425 mg/mL. (d) Device statistics of the preliminary devices shown in (c). The EA additive benefits $J_{sc}$ and $V_{oc}$, resulting in a significant enhancement in performance. (e) Se film processed with 4% EA. The film is initially highly uniform but delaminated after annealing at 200°C; this delamination occurred only for films processed with >3% EA. EA concentration is therefore maintained at 2%; the concentration is maintained at 465 mg/mL for Se. The substrates in (c) and (e) are all 25 mm x 25 mm.

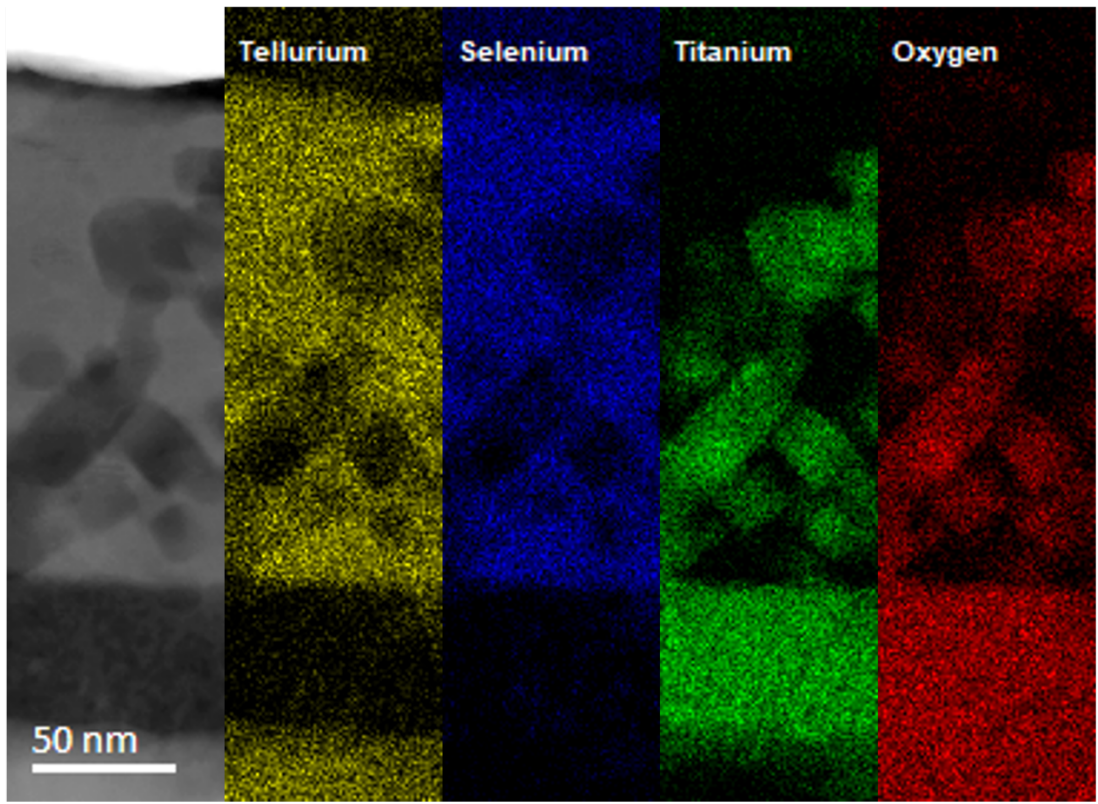

**Figure S3. Cross-sectional STEM EDX Analysis.** X-STEM image and EDS maps of Te, Se, Ti, and O for the 30% Te film.

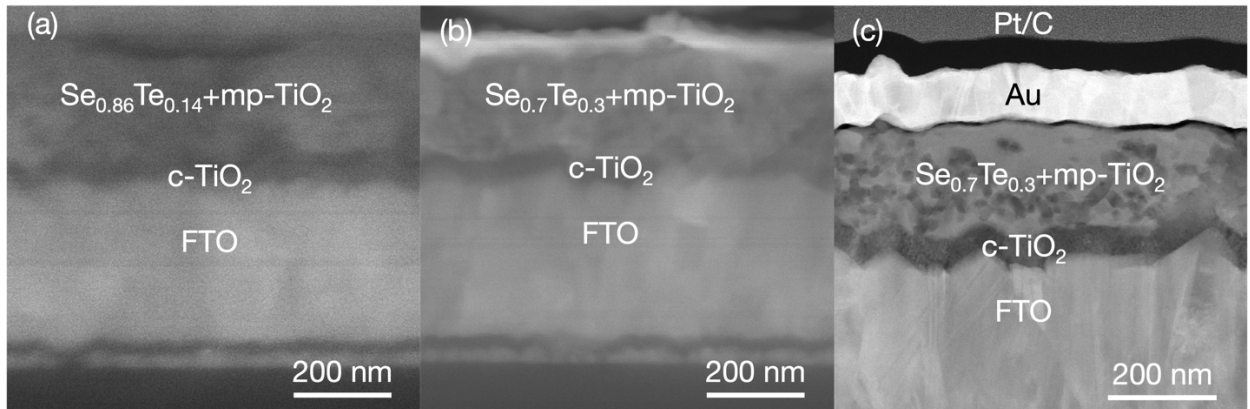

**Figure S4. Cross sectional SEM and HAADF STEM images of Se$_{1-x}$Te$_x$.** XSEM images of (a) 14% Te and (b) 30% Te films, indicating absorber layer thicknesses of approximately 200 nm. (c) High-Angle Annular Dark Field Scanning Transmission Electron Microscopy (HAADF-STEM) image of a 30% Te device, further confirming the thickness and showing the nanostructure of the mp-TiO$_2$ infiltrated absorber. The ultrathin MoO$_x$ layer is present in all images but is not labeled in the images for readability.

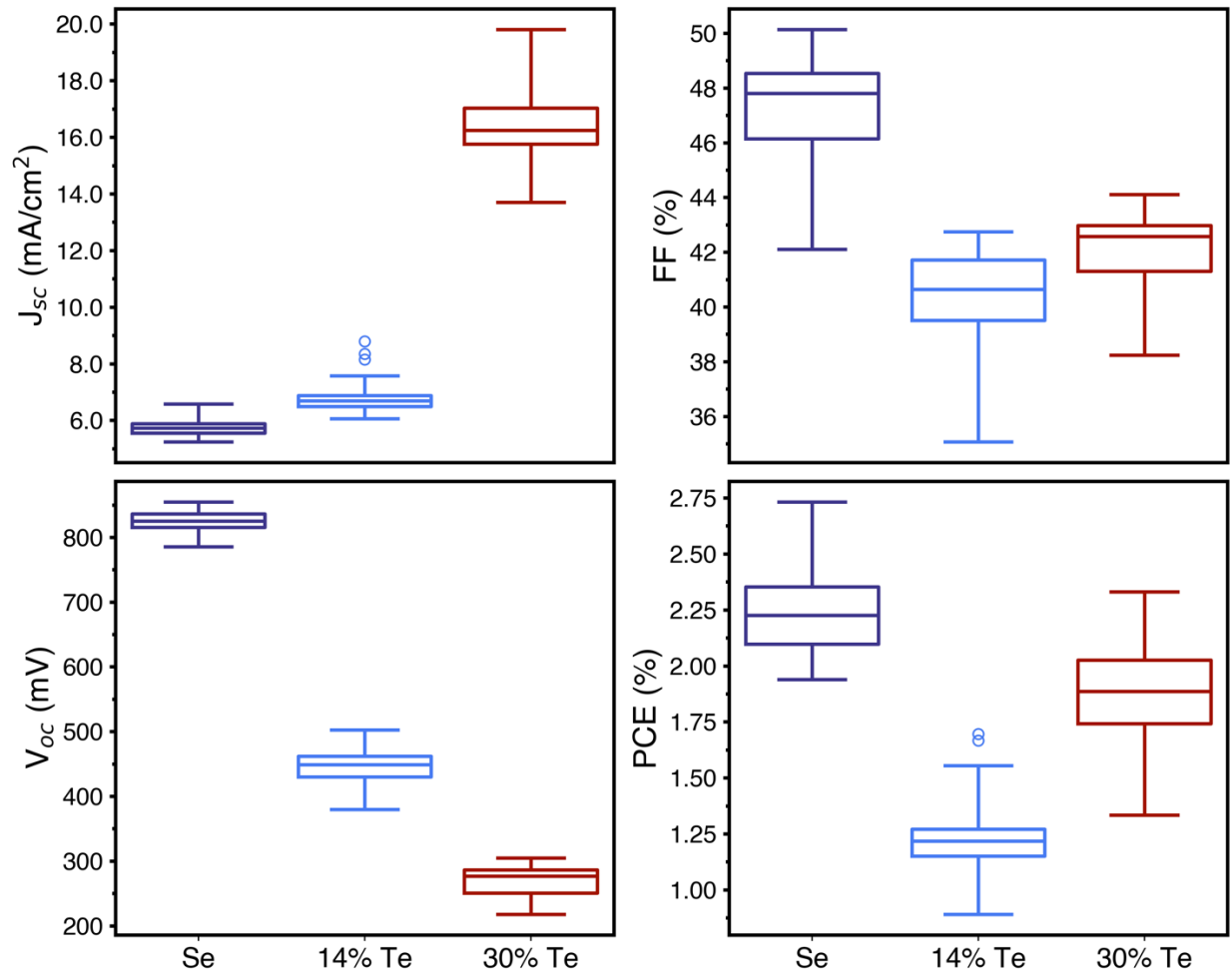

**Figure S5. Device statistics.**

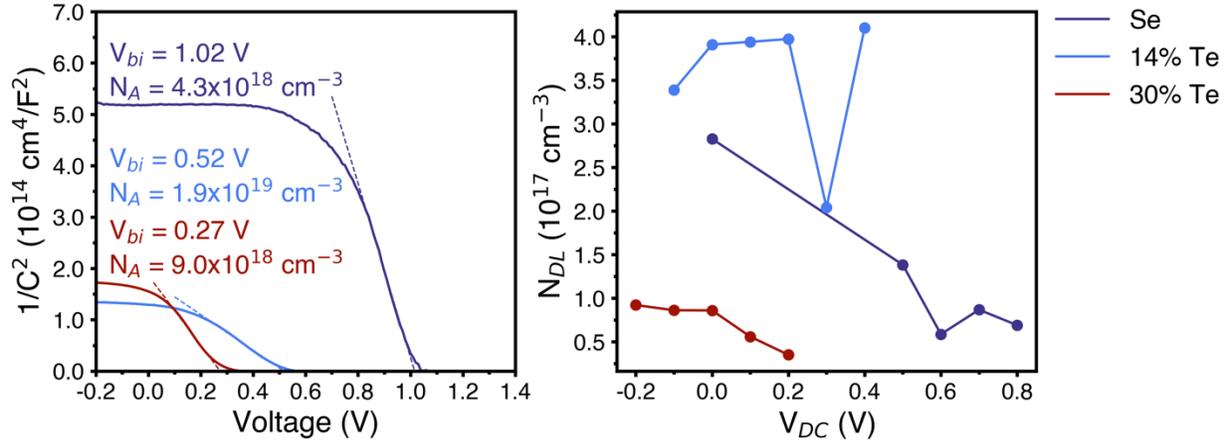

**Figure S6. Capacitance Analysis.** (a) Mott-Schottky plots of $1/C^2$ vs. voltage. The saturation at zero applied bias indicates that these devices are fully depleted. The built-in voltages ($V_{bi}$) are extracted and decrease with added Te content: 1.02V for Se, 0.52V for 14% Te, and 0.27V for 30% Te. The extracted defect concentration, $N_A$, is high ($10^{18}$-$10^{19}$ cm$^{-3}$). (b) Drive level defect concentration, $N_{DL}$, extracted from drive level capacitance measurements. Note that while this is not true drive level capacitance profiling (DLCP) – which is inhibited by the mesoscopic structure – $N_{DL}$ can still be effectively extracted from the equation $N_{DL} = \frac{C_0^3}{q\epsilon A^2 C_1}$ where $C_0$ and $C_1$ are extracted from quadratic fits to the capacitance as a function of signal amplitude. This enables the contribution from interface states to be minimized. By comparing the $N_A$ from C-V and $N_{DL}$, we clearly see that the dominant defects are interface states, with a C-V defect density on the order of $10^{18}$-$10^{19}$ cm$^{-3}$, while the bulk defect density is on the order of $10^{16}$-$10^{17}$ cm$^{-3}$.

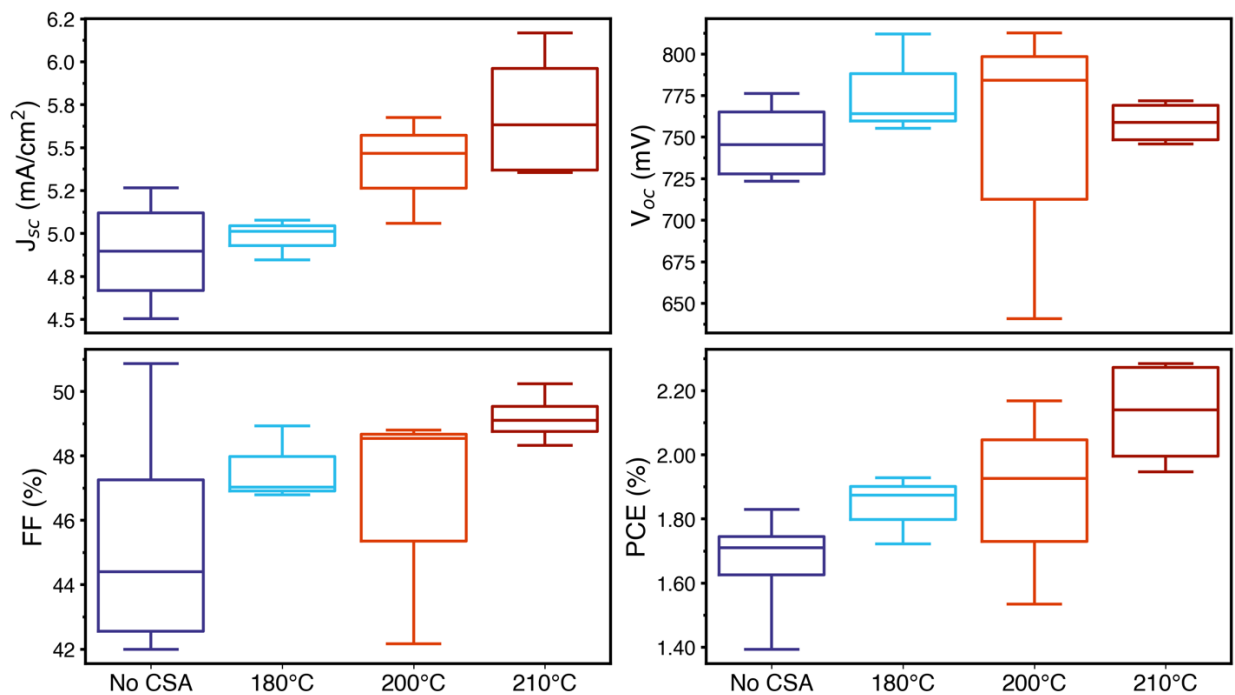

**Figure S7. Effect of post-anneal.** As-fabricated devices were annealed at 180°C, 200°C, and 210°C for 2 min in air after electrode deposition. As in a previous work[1], the post-anneal was found to be beneficial. Therefore, the Se devices undergo a post-anneal.

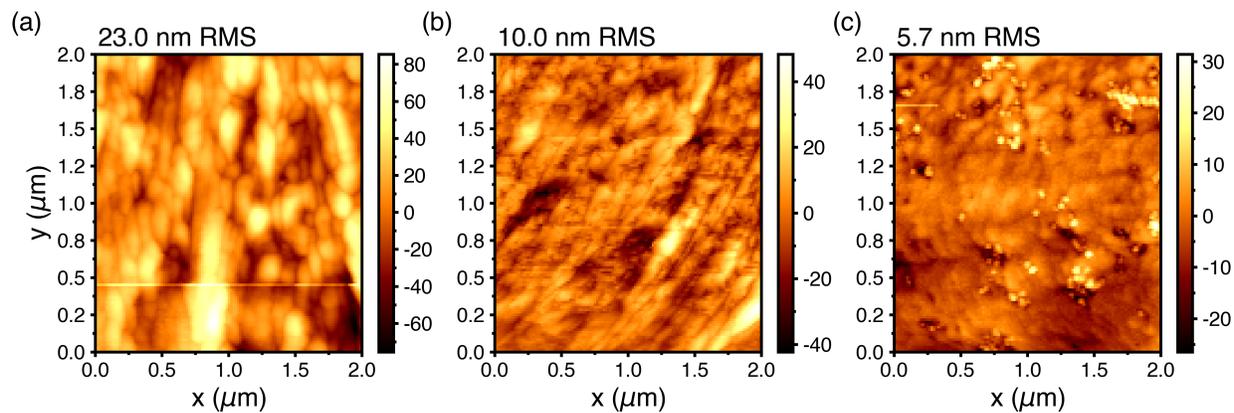

**Figure S8. Atomic force micrographs** of (a) Se, (b) 14% Te, and (c) 30% Te films. The RMS roughness for each is listed. The 30% shows some of the mp-TiO$_2$ poking through but still exhibits a 5.7 nm RMS roughness.